\documentclass[prc,superscriptaddress,noshowpacs,unsortedaddress,twocolumn,showpacs,preprintnumbers,amsmath,amssymb]{revtex4-2}
\usepackage[dvipdfmx]{graphicx}
\usepackage{amsmath,amssymb,times}
\usepackage{color}
\usepackage{ulem}
\usepackage{bm}
\usepackage{here}

\def\gsim{~\,\makebox(1,1){$\stackrel{>}{\widetilde{}}$}\,~}
\def\lsim{~\,\makebox(1,1){$\stackrel{<}{\widetilde{}}$}\,~}

\newcommand{\beq}{\begin{equation}}
\newcommand{\eeq}{\end{equation}}
\newcommand{\bea}{\begin{eqnarray}}
\newcommand{\eea}{\end{eqnarray}}

\newcommand{\bfi}[1]{\mbox{\boldmath $#1$}}

\newcommand{\vK}{{\bfi K}}

\newcommand{\vs}{{\bfi s}}

\newcommand{\vrr}{{\bfi r}}
\newcommand{\vR}{{\bfi R}}

\def\a{\alpha}
\def\b{\beta}

\begin{document}

\title{Neutron skin thickness for $^{208}$Pb from total cross sections of 
neutron  scattering at 14.137~MeV and neutron skin thickness  for  $^{48}$Ca, O, N, C isotopes from reaction and interaction cross sections}

\author{Shingo Tagami}
\affiliation{Department of Physics, Kyushu University, Fukuoka 819-0395, Japan}
%\email[]{sh.tagami@gmail.com}

\author{Takayuki Myo}%
\affiliation{%
 General Education, Faculty of Engineering, Osaka Institute of Technology,
 Osaka 535-8585, Japan
}
%myo@rcnp.osaka-u.ac.jp         

\author{Masanobu Yahiro}
\email[]{orion093g@gmail.com}
\affiliation{Department of Physics, Kyushu University, Fukuoka 819-0395, Japan}             

\begin{abstract}
\begin{description}
\item[Background]
Foster {\it et al.} measured total neutron cross sections  $\sigma_{\rm T}$ of 
n+$^{208}$Pb scattering at  $E_{\rm lab}= 14.137$~MeV, where $E_{\rm lab}$ is an
incident energy per nucleon. 
Carlson {\it et al.} measured  reaction cross sections 
$\sigma_{\rm R}$ for $p$+$^{48}$Ca scattering in $E_{\rm lab}=23 \text{--} 48$~MeV.
Tanaka {\it et al.} measured  interaction cross sections 
$\sigma_{\rm I}$ for $^{42\text{--}51}$Ca + $^{12}$C scattering at 280~MeV/nucleon. 
Ozawa {\it et al.} accumulated the measured values of $\sigma_{\rm I}$ and the extracted matter radii $r_{\rm m}$ 
from $^{4}$He to $^{32}$Mg. 
Bagchi {\it et al.} measured the charge-changing (CC) cross sections and determined proton radii $r_{\rm p}({\rm CC})$ for $^{14,15,17 \text{--} 22}$N 
from the CC cross sections, 
 and found a signal of $^{22}$N being a halo nucleus. 
 Kanungo {\it et al.}  measured the CC cross sections and extracted 
 $r_{\rm p}({\rm CC})$ for $^{12\text{--} 19}$C. 
Kaur {\it et al.} measured the CC cross sections and determined 
$r_{\rm p}({\rm CC})$ for $^{16,18 \text{--} 24}$O. 
\item[Purposes] 
Our 1st aim is to extract neutron tkin thickness $r_{\rm skin}^{208}$ from the 
the $\sigma_{\rm T}$ of n+$^{208}$Pb scattering at  
$E_{\rm lab} =14.137$~MeV.
Our 2nd aim is to  determine $r_{\rm skin}^{48}({\rm skin})$ from $\sigma_{\rm R}$ on p+$^{48}$Ca scattering in $E_{\rm lab}=23 \text{--} 48$~MeV. 
Our  3rd aim is to find light stable nuclei having nuclei having large  $r_{\rm skin}$. 
\item[Methods] 
We use the Kyushu $g$-matrix folding model for lower $E_{\rm lab}$  and the  folding model based on the Love-Franey $t$-matrix for higher $E_{\rm lab}$. We tested the Kyushu (chira)l $g$-matrix folding model for reaction cross sections $\sigma_{\rm R}$ on p+$^{208}$Pb scattering in $20 \lsim E_{\rm lab} \lsim 180$~MeV and found that our folding model is reliable.
As for $^{12}$C+$^{12}$C scattering, 
we tested the Kyushu $g$-matrix folding model and found that the  folding model is reliable for $\sigma_{\rm R}$ in $30  \lsim E_{\rm lab} \lsim 100 $~MeV/u and $250  \lsim E_{\rm lab} \lsim 400 $~MeV/u. 

\item[Results] 
We extract $r_{\rm skin}^{208}=0.309 \pm  0.057$~fm from the $\sigma_{\rm T}$. 
As for $^{48}$Ca, we determine 
$r_{\rm skin}^{48}({\rm skin})=0.163 \pm 0.037~{\rm fm}$ from the 
$\sigma_{\rm R}$ on p+$^{48}$Ca scattering, using the Kyushu $g$-matrix folding model with the D1M-GHFB+AMP proton and neutron densities, where D1M-GHFB+AMP is Gogny-D1M HFB (D1M-GHFB) with angular momentum projection (AMP). We show that D1M-GHFB+AMP is better than D1S-GHFB+AMP for the matter radius and the binding  energy. 
Our skin value is consistent with  $r_{\rm skin}^{48}({\rm CREX})$. 
As for C, N, O isotopes, we find that 
$r_{\rm skin}= 0.267 \pm 0.056$~fm for $^{14}$N and 
$r_{\rm skin}= 0.197 \pm 0.067$~fm for $^{17}$O. 
\item[Conclusion]
The value  $r_{\rm skin}^{208}=0.309 \pm  0.057$~fm agrees with  $r_{\rm skin}^{208}({\rm PREX2})$.
\end{description}
 \end{abstract}

\maketitle

\section{Introduction}
\label{Sec:Introduction}

Horowitz {\it et al.} \cite{PRC.63.025501} proposed a direct measurement 
for neutron skin thickness $r_{\rm skin}$ = $r_{\rm n} - r_{\rm p}$, where 
$r_{\rm n}$ and $r_{\rm p}$ are the root-mean-square radii of neutrons  
and protons, respectively. 

The PREX collaboration has reported~\cite{Adhikari:2021phr} 
\begin{equation}
r_{\rm skin}^{208}({\rm PREX2}) = 0.283\pm 0.071= 0.212 \text{--} 0.354
\,{\rm fm}.
\label{Eq:Experimental constraint 208}
\end{equation}
The CREX group has presented~\cite{CREX:2022kgg} 
\bea
r_{\rm skin}^{48}({\rm CREX})=0.121 \pm 0.026\ {\rm (exp)} \pm 0.024\ {\rm (model)}\,{\rm fm}. 
 \label{CREX-value}
 \eea
The PREX2 and CREX values are reliable for $^{208}$Pb and $^{48}$Ca, respectively. 
Using  the $r_{\rm p}$ deduced from the electron scattering of Refs.~\cite{Angeli:2013epw,Brown:2013mga}, one can obtain $r_{\rm n}$ and matter radii $r_{\rm m}$ for 
PREX2 and CREX, as shown in Table   \ref{TW-prex}. 

%\squeezetable
%%%%%%%%%%%%%%
\begin{table}[htb]
\begin{center}
\caption
{Values of   $r_{\rm m}({\rm exp})$,  $r_{\rm n}({\rm exp})$, 
$r_{\rm skin}({\rm exp})$ 
together with $r_{\rm p}({\rm exp})$ 
deduced from the electron scattering~\cite{Angeli:2013epw,Brown:2013mga}. 
The radii are shown in units of fm.  
 }
\begin{tabular}{cccccc}
\hline\hline
 & . & $r_{\rm p}({\rm exp})$ & $r_{\rm m}({\rm exp})$ &  $r_{\rm n}({\rm exp})$ & $r_{\rm skin}({\rm exp})$ \\
\hline
  & PREX2 & $5.444$ & $5.617 \pm	0.044$ & $5.727 \pm 0.071$ & $0.283	\pm 0.071$ \\
  & CREX & $3.385$ & $3.456 \pm 0.030$ & $3.506 \pm 0.050$ & $0.121 \pm 0.050$ \\
\hline
\end{tabular}
 \label{TW-prex}
 \end{center} 
 \end{table}
%%%%%%%%%%%%%% 

In Ref.~\cite{Tagami:2020bee}, we extracted 
$r_{\rm skin}^{208}=0.278 \pm 0.035$~fm from  
reaction cross sections $\sigma_{\rm R}$ on p+$^{208}$Pb scattering, using the chiral (Kyushu) $g$-matrix folding model~\cite{Toyokawa:2017pdd,PRC.101.014620} with the densities calculated with Gogny-D1S HFB (D1S-GHFB) with angular momentum projection (AMP).  
The central value almost agrees with that of $r_{\rm skin}^{208}({\rm PREX2})$.
In this paper, we try to find nucleus having thick skin-value 
in order to support $r_{\rm skin}^{208}({\rm PREX2})$. 

In Ref.~\cite{Wakasa:2022ite}, we tested the Kyushu $g$-matrix folding model for $\sigma_{\rm R}$ on p+$^{208}$Pb scattering in $20 \lsim E_{\rm lab} \lsim 180$~MeV and find that our folding model is reliable there, 
using the $\sigma_{\rm R}({\rm PREX2})$ calculated 
with the folding model with the neutron density scaled to 
$r_{\rm n}^{208}({\rm PREX2})$; note that 
the $r_{\rm p}^{208}$ calculated with D1S-GHFB+AMP agrees with 
the $r_{\rm p}^{208}({\rm PREX2})$ of Ref.~~\cite{Brown:2013mga}
In the paper, we extracted  $r_{\rm skin}$ and $r_{\rm m}$ from the data 
$\sigma_{\rm R}$ for $^{208}$Pb, $^{58}$Ni, $^{48,40}$Ca , $^{12}$C targets, using the Kyushu $g$-matrix folding model with the densities calculated with D1S-GHFB+AMP. As a way of a fine-tuning factor $f$, we proposed 
the ESP-f (experimental scaling procedure). The  ESP-f is a reliable way for $^{208}$Pb, $^{58}$Ni, $^{40,48}$Ca, $^{12}$C. 
As for $^{12}$C, the ESP-f is nothing but a method from interaction 
cross sections $r_{\rm m}(\sigma_{\rm I})$ to $r_{\rm m}(\sigma_{\rm R})$. 

As for $^{4}$He scattering, we determined $r_{\rm skin}^{208}=$ 0.416$\pm$0.146 fm for  $^{208}$Pb in 30 $\lsim E_{\rm lab}\lsim$ 50 MeV/u with the Kyushu $g$-matrix folding model with the D1S-GHFB+AMP proton and neutron densities, and extracted $r_{\rm skin}$ and 
$r_{\rm m}$ for $^{40}$Ca and Sn isotopes~\cite{Matsuzaki:2021hdm}.   
In this paper, we test the Kyushu $g$-matrix folding model for $\sigma_{\rm R}$ on $^{4}$He+$^{208}$Pb scattering in $30 \lsim E_{\rm lab} \lsim 180$~MeV with the D1S-GHFB+AMP proton and neutron densities, using the ESP-f, that is, we extracted $r_{\rm skin}^{208}$ as 
the $\sigma_{\rm R}({\rm PREX2})$ as a reference (reliable) value.  

Foster {\it et al.} measured total neutron cross sections  $\sigma_{\rm T}$ of n+$^{208}$Pb scattering at  
$E_{\rm lab}= 14.137$~MeV~\cite{PhysRevC.3.576}. Our aim is to extract $r_{\rm skin}^{208}$ from the the $\sigma_{\rm T}$

The measured values of  $\sigma_{\rm I}$ and the extracted matter radii $r_{\rm m}(\sigma_{\rm I})$ 
are accumulated  from $^{4}$He to $^{32}$Mg in Ref.~\cite{Ozawa:2001hb}; see Table 1 of Ref.~\cite{Ozawa:2001hb} for the $\sigma_{\rm I}$ and see Table 3 (Glauber model in the optical limit) for the $r_{\rm m}(\sigma_{\rm I})$.

Bagchi {\it et al.} measured charge-changing (CC) cross sections 
 around 900 MeV/u at GSI and determined 
 the proton radii $r_{\rm p}({\rm CC})$ for  $^{14,15,17 \text{--} 22}$N~\cite{Bagchi:2019sua}. 
As for  $^{14,15}$N, the $r_{\rm p}({\rm e^{-}})$  are obtainable 
with the isotope shift based on electron scattering~\cite{Angeli:2013epw}. 
As for $A=14,15$, the $r_{\rm p}({\rm CC}) $  are consistent with 
those of electron scattering.  
They extracted matter radii $r_{\rm m}$ and neutron skin thickness 
$r_{\rm skin}$ from interaction cross sections 
$\sigma_{\rm I}$~\cite{Ozawa:2001hb} for  $^{14,15,17 \text{--} 22}$N, using  the
Glauber model where the $r_{\rm p}({\rm CC})$ were used and 
the neutron radius $r_{\rm n}$ were evaluated with the harmonic oscillator density. 
They mentioned that an increase in  $r_{\rm m}$ from $^{21}$N to $^{22}$N is a signal of $^{22}$N being a halo nucleus. 

Kaur {\it et al.} measured charge-changing (CC) cross sections 
and  $\sigma_{\rm I}$ for $^{16, 18 \text{--} 24}$O+$^{12}$C scattering at 
 around 900$A$ MeV and presented the values of  
 the $r_{\rm p}({\rm CC})$ and the $r_{\rm m}$
 for  $^{16, 18 \text{--} 24}$O, using  the
Glauber model~\cite{Kaur:2022yoh}; see Table I of 
 Ref.~\cite{Kaur:2022yoh}.  
As for $A=16,18$, the $r_{\rm p}({\rm CC}) $  are consistent with 
the $r_{\rm p}({\rm e^{-}})$.  
They also extracted $r_{\rm skin}$ from the $r_{\rm p}({\rm CC})$ and 
the $r_{\rm m}$ 
for  $^{16, 18 \text{--} 24}$O. 

Kanungo {\it et al.} measured the CC cross sections for $^{12,14 \text{--} 19}$C+$^{12}$C scattering to determine the  $r_{\rm p}({\rm CC}) $~\cite{Kanungo:2016tmz}. Their  values are consistent with those of electron scattering for $A=12,14$.  

Dobrovolsky  {\it et al.} measured the absolute differential cross sections for small-angle elastic scattering of on $^{12,14 \text{--} 17}$C on a proton target at energies near 700 MeV/u  and 
determined  $r_{\rm m}$ for $^{12,14 \text{--} 17}$C 
by using the Glauber model.~\cite{Dobrovolsky:2021ggf}. 
Using  $r_{\rm p}({\rm e^{-}})$ for $^{12,14}$C and  the $r_{\rm p}({\rm CC})$ of Ref.~\cite{Kanungo:2016tmz} for $^{15,16,17}$C, they extracted 
$r_{\rm skin}$; see Table 2 of Ref.~\cite{Dobrovolsky:2021ggf}.

Reaction cross sections $\sigma_{\rm R}$, interaction cross sections 
$\sigma_{\rm I}$ for charged projectiles and $\sigma_{\rm I}$ for neutron are  standard observable of determining $r_{\rm skin}$, when the  $r_{\rm p}({\rm exp})$ is calculated with  the isotope shift method based on the electron scattering~\cite{Angeli:2013epw}. 
Good data on $\sigma_{\rm R}$  for $p$+$^{48}$Ca scattering
are available in  Ref.~\cite{Carlson:1994fq}.

As for $^{12}$C+$^{12}$C scattering, 
we tested the Kyushu $g$-matrix folding model and 
found that the folding model~\cite{Toyokawa:2017pdd} is reliable 
for $\sigma_{\rm R}$ in $30  \lsim E_{\rm lab} \lsim 100 $~MeV and $250  \lsim E_{\rm lab} \lsim 400 $~MeV~\cite{Tagami:2019svt}.

Tanaka {\it et al} measured  interaction cross sections 
$\sigma_{\rm I}$ for $^{42\text{--}51}$Ca + $^{12}$C scattering at 280~MeV/nucleon and determined 
$r_{\rm skin}$ for $^{42\text{--}51}$Ca~\cite{Tanaka:2019pdo}. 
As for $^{48}$Ca, Tanaka {\it et al.} extracted 
$r_{\rm skin}^{48}=0.146 \pm 0.06$~fm,  using the Glauber model (the 
optical limit) with the Woods-Saxon proton and neutron densities.

In Ref.~\cite{TAKECHI2021104923}, we reanalyzed  the data, using the Kyushu $g$-matrix folding model with the D1S-GHFB densities for $^{43,45,47,49,51}$Ca and the D1S-GHFB+AMP densities for $^{42,44,46,48,50}$Ca. Their skin values almost agree with ours, except for $^{48}$Ca.  
D1S is thus reliable for $^{42\text{--}47, 49\text{--}51}$Ca. 
There is non-negligible difference between  our value $r_{\rm skin}^{48}=0.105 \pm 0.06$~fm and theirs $r_{\rm skin}^{48}=0.146 \pm 0.06$~fm. The fact indicates that we should carefully choose  proton and neutron densities   for $^{48}$Ca. 

Only as for $^{48}$Ca, we then choose D1M~\cite{Gonzalez-Boquera:2017rzy} in stead of D1S, since the D1M-GHFB+AMP calculation yields better agreement with the total energy than the D1S-GHFB+AMP one~\cite{TAGAMI2022105155}.  
The further reason why we take D1M is shown in Sec.~\ref{Comparison between D1S and D1M}.

We determined $r_{\rm m}(\sigma_{\rm R})$ for $^{12}$C, as shown in 
Table \ref{reference values-0}. 
Our result $r_{\rm m}(\sigma_{\rm R})$ of Ref.~\cite{Wakasa:2022ite} based on 
p+ $^{12}$C scattering almost agree with that of Ref.~\cite{TAGAMI2023106675} based on $^{12}$C+$^{12}$C scattering each other. 
Note that there is no fine-tuning factor for $^{12}$C+$^{12}$C scattering 
and we used the ESP-f for p+$^{12}$C scattering. 

%\squeezetable
%%%%%%%%%%%%%%
\begin{table}[htb]
\begin{center}
\caption
{Values of   $r_{\rm m}(\sigma_{\rm R})$ and $r_{\rm m}(\sigma_{\rm I})$. 
The two values $r_{\rm m}(\sigma_{\rm I})$ are taken from 
the accumulation paper of  Ref.~\cite{Ozawa:2001hb}. 
 The first $r_{\rm m}(\sigma_{\rm R})$ are determined from 
 the $\sigma_{\rm R}$ of p scattering in Ref.~\cite{Wakasa:2022ite}, whereas the 2nd $r_{\rm m}(\sigma_{\rm R})$ is extracted from 
 the $\sigma_{\rm R}$ of  $^{12}$C+$^{12}$C scattering in Ref.~\cite{TAGAMI2023106675}. 
The radii are shown in units of fm.  
 }
\begin{tabular}{cccccc}
\hline\hline
 & $r_{\rm m}(\sigma_{\rm R})$ & $r_{\rm m}(\sigma_{\rm R})$ & 
 $r_{\rm m}(\sigma_{\rm I})$ &  $r_{\rm m}(\sigma_{\rm I})$ & \\
\hline
 $^{12}$C & $2.340 	\pm 0.009$ & $2.352 \pm 0.013$ &$2.31 \pm 0.02$  & $2.35 \pm 0.02$  \\
\hline
\end{tabular}
 \label{reference values-0}
 \end{center} 
 \end{table}
%%%%%%%%%%%%%% 

The chiral nucleon-nucleon (NN) forces used in the Kyushu $g$-matrix folding model has a cutoff of  550 MeV. For this reason, the model is applicable for $E_{\rm lab} \lsim 410$~MeV.  
At $E_{\rm lab} \gsim 500 $~MeV, in fact,  
we extracted $r_{\rm skin}$ and $r_{\rm m}$  
for $^{208}$Pb by using the folding model~\cite{WAKASA2022106101}  based on the Love-Franey (LF) $t$-matrix~\cite{LF} model.  
Our $r_{\rm skin}$ values, $0.325 \pm 0.076$~fm  for D1S+AMP and 
$0.333  \pm 0.076$~fm  for D1M+AMP, are consistent with 
$r_{\rm skin}^{208}({\rm PREX2})$. Note that the former value is 
very close to the latter. 

Our 1st aim is to extract $r_{\rm skin}^{208}$ from the 
the $\sigma_{\rm T}$ of Ref.~\cite{PhysRevC.3.576} of n+$^{208}$Pb scattering at  $E_{\rm lab} =14.137$~MeV.
Our 2nd aim is to  determine $r_{\rm skin}^{48}({\rm skin})$ from the $\sigma_{\rm R}$ of Ref.~\cite{Carlson:1994fq} on p+$^{48}$Ca scattering in $E_{\rm lab}=23 \text{--} 48$~MeV. 
Our  3rd aim is to find light stable nuclei having nuclei with large  $r_{\rm skin}$.

Our model is formulated in Sec.~\ref{Sec:Method}, 
and our results is shown in Sec.~\ref{Results}.  
Section \ref{Sec:Summary} is devoted to a summary.

\section{Mehod}
\label{Sec:Method}

The $g$-matrix folding model~\cite{Brieva-Rook,Amos,CEG07,Minomo:2011bb, Sumi:2012fr, Egashira:2014zda,Watanabe:2014zea,Toyokawa:2014yma,Toyokawa:2015zxa,Toyokawa:2017pdd} is a standard way of 
determining $r_{\rm skin}$ and/or $r_{\rm m}$ from $\sigma_{\rm R}$ and 
 $\sigma_{\rm I}$.
In the model, the  potential is obtained by folding  the $g$-matrix with projectile and target densities.

Applying the folding model based on 
the Melbourne $g$-matrix~\cite{Amos}   for  $\sigma_{\rm R}$ of Mg isotopes, we 
deduced the $r_{\rm m}$ for   
Mg isotopes~\cite{Watanabe:2014zea}, and discovered 
that $^{31}$Ne is a halo nucleus with large deformation~\cite{Minomo:2011bb}. 

Kohno calculated the $g$ matrix  for the symmetric nuclear matter, 
using the Brueckner-Hartree-Fock method with chiral N$^{3}$LO 2NFs and NNLO 3NFs~\cite{Koh13}. 
He set $c_D=-2.5$ and $c_E=0.25$ so that  the energy per nucleon can  become minimum 
at $\rho = \rho_{0}$~\cite{Toyokawa:2017pdd}.

Toyokawa {\it et al.} localized the non-local chiral  $g$ matrix 
into three-range Gaussian forms by using the localization method proposed 
by the Melbourne group~\cite{von-Geramb-1991,Amos-1994,Amos}. 
The resulting local  $g$ matrix is called  ``Kyushu  $g$-matrix''. 

The  Kyushu $g$-matrix folding model is successful in reproducing the differential cross sections $d\sigma/d\Omega$ and the vector analyzing power $A_y$
for polarized  proton scattering  on various targets at $E_{\rm lab}=65$~MeV~\cite{Toyokawa:2014yma}, 
and $d\sigma/d\Omega$ for $^4$He scattering at $E_{\rm lab}=72$~MeV per nucleon~\cite{Toyokawa:2015zxa}. 
This is true for $\sigma_{\rm R}$ of $^4$He scattering 
in $E_{\rm lab}=30 \sim 200$~MeV per nucleon~\cite{Toyokawa:2017pdd}. 

In this paper, we use the Kyushu $g$-matrix  folding model~\cite{Toyokawa:2017pdd} for lower energies 
and the LF folding model for higher energies.

In the the Kyushu $g$-matrix folding model, the potential $U$ consists 
of the direct part ($U^{\rm DR}$) and the exchange part ($U^{\rm EX}$) 
defined by
\bea
\label{eq:UD}
U^{\rm DR}(\vR) \hspace*{-0.15cm} &=& \hspace*{-0.15cm} 
\sum_{\mu,\nu}\int \rho^{\mu}_{\rm P}(\vrr_{\rm P}) 
            \rho^{\nu}_{\rm T}(\vrr_{\rm T})
            g^{\rm DR}_{\mu\nu}(s) d \vrr_{\rm P} d \vrr_{\rm T}, \\
\label{eq:UEX}
U^{\rm EX}(\vR) \hspace*{-0.15cm} &=& \hspace*{-0.15cm}\sum_{\mu,\nu} 
\int \rho^{\mu}_{\rm P}(\vrr_{\rm P},\vrr_{\rm P}-\vs)
\rho^{\nu}_{\rm T}(\vrr_{\rm T},\vrr_{\rm T}+\vs) \nonumber \\
            &&~~\hspace*{-0.5cm}\times g^{\rm EX}_{\mu\nu}(s) \exp{[-i\vK(\vR) \cdot \vs/M]}
            d \vrr_{\rm P} d \vrr_{\rm T},~~~~
            \label{U-EX}
\eea
where $\vs=\vrr_{\rm P}-\vrr_{\rm T}+\vR$ 
for the coordinate $\vR$ between a projectile (P) and a target (T). The coordinate 
$\vrr_{\rm P}$ 
($\vrr_{\rm T}$) denotes the location for the interacting nucleon 
measured from the center-of-mass of P (T) and $M=A A_{\rm T}/(A +A_{\rm T})$
for the mass number $A$ ($A_{\rm T}$) of P (T). 
Each of $\mu$ and $\nu$ stands for the $z$-component
of isospin; 1/2 means neutron and $-$1/2 does proton.

The direct and exchange parts, $g^{\rm DR}_{\mu\nu}$ and 
$g^{\rm EX}_{\mu\nu}$, of the $g$ matrix are described by
\begin{align}
g_{\mu\nu}^{\rm DR}(s) 
&=
\displaystyle{\frac{1}{4} \sum_S} \hat{S}^2 g_{\mu\nu}^{S1}
 (s) \hspace*{0.1cm}  \hspace*{0.1cm} 
 {\rm for} \hspace*{0.1cm} \mu+\nu = \pm 1,
 \\
g_{\mu\nu}^{\rm DR}(s) 
&=
\displaystyle{\frac{1}{8} \sum_{S,T}} 
\hat{S}^2 g_{\mu\nu}^{ST}(s) 
\hspace*{0.1cm}  \hspace*{0.1cm} 
{\rm for} \hspace*{0.1cm} \mu+\nu = 0,
\\
g_{\mu\nu}^{\rm EX}(s) 
&=
\displaystyle{\frac{1}{4} \sum_S} (-1)^{S+1} 
\hat{S}^2 g_{\mu\nu}^{S1} (s) 
\hspace*{0.1cm}  \hspace*{0.1cm} 
{\rm for} \hspace*{0.1cm} \mu+\nu = \pm 1, 
\\
g_{\mu\nu}^{\rm EX}(s) 
&=
\displaystyle{\frac{1}{8} \sum_{S,T}} (-1)^{S+T} 
\hat{S}^2 g_{\mu\nu}^{ST}(s) 
\hspace*{0.1cm}  \hspace*{0.1cm}
{\rm for} \hspace*{0.1cm} \mu+\nu = 0 
,
\end{align}
where $\hat{S} = {\sqrt {2S+1}}$ and $g_{\mu\nu}^{ST}$ are 
the spin-isospin components of the $g$-matrix interaction.
The Kyushu $g$-matrix~\cite{Toyokawa:2017pdd} is constructed from the chiral 2NFs and 3NFs interaction with the cutoff 550~MeV.  
In the LF $t$-matrix folding model, the chiral $g$-matrix is placed by the LF $t$-matrix.~\cite{LF} 
The formulation for proton+nucleus scattering is shown in 
Ref.~\cite {Egashira:2014zda}.

As for a  $^{12}$C target, we use the phenomenological density of Ref.~\cite{C12-density}. 
We use the Gogny-D1S Hartree-Fock-Bogoliubov 
(D1S-GHFB)  method for O-isotope densities, since 
effects of angular momentum projection (AMP) are negligible for 
spherical nuclei such as O isotopes. 
As  C isotopes, we use  the D1S+GHFB+AMP densities 
for $A=14,16,18$ and the phenomenological one for $A=12$.  
As odd nuclei such as N isotopes, 
the D1S+GHFB+AMP method is not feasible; 
this point is explained in Ref.~\cite{Tagami:2019svt}. 
We use SLy7~\cite{Chabanat:1997un,Schunck:2016uvm,Matsuzaki:2021hdm} that is an improved version of SLy4. The SLy7 were used for $^{4}$He+$^{208}$Pb scattering in order to extract $r_{\rm skin}^{208}=$ 0.416$\pm$0.146 fm~\cite{Matsuzaki:2021hdm}. 
As for $^{208}$Pb, the SLy7 yields the same $r_{\rm m}$ as D1S-GHFB+AMP. 
This is true for N isotopes. 

The scaled density $\rho_{\rm scaling}(\vrr)$ is obtained 
from the original projectile density $\rho(\vrr)$ as
\bea
\rho_{\rm scaling}(\vrr)=\frac{1}{\a^3}\rho(\vrr/\a)
\eea
with a scaling factor
\bea
\a=\sqrt{ \frac{\langle \vrr^2 \rangle_{\rm scaling}}{\langle \vrr^2 \rangle}} .
\eea

In order to extract the $r_{\rm m}$ from the measured 
$\sigma_{\rm I}$ and $\sigma_{\rm R}$, 
we scale the proton and neutron densities 
so as to reproduce the $\sigma_{\rm I}$ 
under the condition 
that  $r_{\rm p,scaling}=r_{\rm p}({\rm exp})$, where 
$r_{\rm p}({\rm exp})$ stands for either $r_{\rm p}({\rm e^{-}})$ 
or $r_{\rm p}({\rm CC}) $; note that $Ar_{\rm m}^2=Zr_{\rm p}^2+Nr_{\rm n}^2$. 

\subsection{Relation between the Kyushu $t$-matrix folding model 
and the LF $t$-matrix folding model}

 Now, we compare the LF $t$-matrix folding model with 
the Kyushu $t$-matrix folding model for ${}^{12}{\rm C}$+${}^{12}{\rm C}$ 
scattering at $30 \lsim E_{\rm lab} \lsim 950$ MeV/u. 
The difference between the results of the Kyushu $g$-matrix folding model and those of the Kyushu $t$-matrix folding model is small at 
$E_{\rm lab} = 372$ MeV. 
Since the chiral  $t$-matrix has a cutoff of 
550~MeV, the  results of the chiral  $t$-matrix folding model are reliable  in  
$E_{\rm lab} \lsim 410$ MeV/u, where we use the phenomenological projectile and target densities for both the models.
 The results of the Kyushu $t$-matrix folding  model agree with 
those of the LF $t$-matrix folding model $F \sigma_R^{\rm LF}$ with $F=0.93766$ at $E_{\rm lab}=410$ MeV/u. 

\begin{figure}[H]
\begin{center}
 \includegraphics[width=0.5\textwidth,clip]{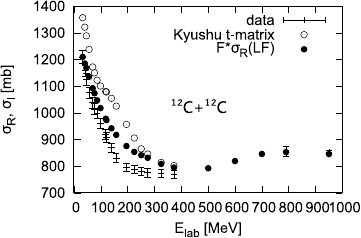}
 \caption{ 
 $E_{\rm lab}$ dependence of $\sigma_{\rm R}$ and $\sigma_{\rm I}$ 
 for $^{12}$C+$^{12}$C scattering. 
  Open circles stand for the results of the  Kyushu $t$-matrix folding  model with the phenomenological projectile and target densities. 
  Closed circles correspond to the  $F \sigma_{\rm R}({\rm LF})$ 
 with $F=0.93766$. 
  The data are taken from Refs.~\cite{Ozawa:2001hb,Tanihata:1988ub,Ozawa:2000gx,Takechi:2009zz}. 
   }
 \label{Fig-RXsec-12C+12C}
\end{center}
\end{figure}

As shown in Fig.~\ref{Fig-RXsec-12C+12C},  
the fine-tuning factor $F$ satisfies  
$\sigma_{\rm I}({\rm exp})=F \sigma_{\rm R}({\rm LF})$ 
for $^{12}$C+$^{12}$C scattering at 790 and 950~MeV/nucleon, 
as shown in Fig. \ref{Fig-RXsec-12C+12C}. 
In $350 \lsim E_{\rm lab} \lsim 400$ MeV/nucleon, 
the results of Kyushu $t$-matrix folding model almost agree with 
$F \sigma_R^{\rm LF}$ with $F=0.93766$, 
where $\sigma_R^{\rm LF}$ is the $\sigma_R$ 
calculated with the LF $t$-matrix folding model. 

The $F$ is used for scattering of C, N, O isotopes on a $^{12}$C target.

%%%%%%%%%%%%%%%%%%%%%%%
%%%  Figure
%%%%%%%%%%%%%%%%%%%%%%%

%Results
\section{Results}
\label{Results} 

\subsection{Determination of $r_{\rm skin}^{208}$ from the 
the total cross sections at $E_{\rm lab}=14.137$~MeV}

\begin{figure}[H]
\begin{center}
 \includegraphics[width=0.5\textwidth,clip]{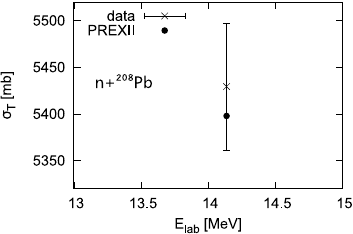}
 \caption{ 
 $E_{\rm lab}$ dependence of $\sigma_{\rm T}$  for n+$^{208}$Pb scattering  
 at $E_{\rm lab}=14.137$~MeV. 
 Closed circles denote the the total cross sections $\sigma_{\rm T}({\rm PREX2})$ calculated 
with the folding model with the neutron density scaled to 
$r_{\rm n}^{208}({\rm PREX2})$, where 
the $r_{\rm p}^{208}$ calculated with D1S-GHFB+AMP agrees with 
the $r_{\rm p}^{208}({\rm PREX2})$ of Ref.~~\cite{Brown:2013mga}
The data are taken from Ref.~\cite{PhysRevC.3.576}. 
   }
 \label{Fig-TXsec-n+Pb}
\end{center}
\end{figure}

Figure~\ref{Fig-TXsec-n+Pb} shows  $E_{\rm lab}$ dependence 
of $\sigma_{\rm T}$ for n+$^{208}$Pb scattering.  
Closed circles denote the the total cross sections $\sigma_{\rm T}({\rm PREX2})$ calculated 
with the folding model with the neutron density scaled to 
$r_{\rm n}^{208}({\rm PREX2})$, where 
the $r_{\rm p}^{208}$ calculated with D1S-GHFB+AMP agrees with 
the $r_{\rm p}^{208}({\rm PREX2})$ of Ref.~~\cite{Brown:2013mga}

%%%%%%%%%%%%%%%%%%%%%%%
%%%  Figure
%%%%%%%%%%%%%%%%%%%%%%%

Scaling the neutron PREX2 density for $^{208}$Pb with 
the ESP-f method to the data~\cite{PhysRevC.3.576} on the total cross sections at 14.137~MeV, 
we can obtain $r_{\rm skin}^{208}=0.309 \pm  0.057$~fm. The value agrees with $r_{\rm skin}^{208}({\rm PREX2})$.

\begin{figure}[H]
\begin{center}
 \includegraphics[width=0.5\textwidth,clip]{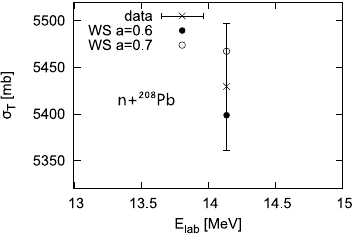}
 \caption{ 
 $E_{\rm lab}$ dependence of $\sigma_{\rm T}$  for n+$^{208}$Pb scattering  
 at $E_{\rm lab}=14.137$~MeV. 
 An open circle stands for the result of 
the Woods-Saxon type neutron density ($r_{\ WS}=6.59$~fm, $a_{\ WS}=0.7$~fm) 
fitted to the central value of PREX2 and the D1S-GHFB+AMP proton density, and  a close circle denotes
the result of the Woods-Saxon type neutron density($r_{\ WS}=6.81$~fm, $a_{\ WS}=0.6$~fm) and the D1S-GHFB+AMP proton density. 
 The data are taken from Ref.~\cite{PhysRevC.3.576}. 
   }
 \label{Fig-TXsec-n+Pb-v2}
\end{center}
\end{figure}

Figure~\ref{Fig-TXsec-n+Pb-v2} shows total cross sections 
 $\sigma_{\rm T}$ of n+$^{208}$Pb scattering as a 
function of  $E_{\rm lab}$.  
An open circle stands for the result of 
the Woods-Saxon type neutron density ($r_{\ WS}=6.59$~fm, $a_{\ WS}=0.7$~fm) 
fitted to the central value of PREX2 and the D1S-GHFB+AMP neutron  density, and  a close circle denotes
the result of the Woods-Saxon type neutron density ($r_{\ WS}=6.81$~fm, $a_{\ WS}=0.6$~fm) and the D1S-GHFB+AMP neutron  density. 
The former (latter) result is near the upper (lower) bound of the data~\cite{PhysRevC.3.576}.  
The central value of the data~\cite{PhysRevC.3.576} indicates   $a_{\ WS} \approx 0.65$~fm.

%%%%%%%%%%%%%%%%%%%%%%%
%%%  Figure
%%%%%%%%%%%%%%%%%%%%%%%

\subsection{Test of the Kyushu $g$-matrix folding model for $^{4}$He+$^{208}$Pb scattering  in $30 \lsim E_{\rm lab} \lsim 180$~MeV}

\begin{figure}[H]
\begin{center}
 \includegraphics[width=0.5\textwidth,clip]{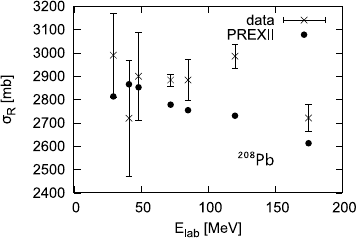}
 \caption{ 
 $E_{\rm lab}$ dependence of $\sigma_{\rm R}$  for $^{4}$He+$^{208}$Pb scattering  in $30 \lsim E_{\rm lab} \lsim 180$~MeV. 
 Closed circles denote the $\sigma_{\rm R}({\rm PREX2})$ calculated 
with the folding model with the neutron density scaled to 
$r_{\rm n}^{208}({\rm PREX2})$, where 
the $r_{\rm p}^{208}$ calculated with D1S-GHFB+AMP agrees with 
the $r_{\rm p}^{208}({\rm PREX2})$ of Ref.~~\cite{Brown:2013mga}
The data are taken from Refs.~\cite{INGEMARSSON20003,BONIN1985381}. 
   }
 \label{Fig-RXsec-He+Pb}
\end{center}
\end{figure}

Figure~\ref{Fig-RXsec-He+Pb} shows  $E_{\rm lab}$ dependence 
of $\sigma_{\rm R}$ for $^{4}$He+$^{208}$Pb scattering.  
Closed circles denote the $\sigma_{\rm R}({\rm PREX2})$ calculated 
with the folding model with the neutron density scaled to 
$r_{\rm n}^{208}({\rm PREX2})$, where 
the $r_{\rm p}^{208}$ calculated with D1S-GHFB+AMP agrees with 
the $r_{\rm p}^{208}({\rm PREX2})$ of Ref.~~\cite{Brown:2013mga}

The $\sigma_{\rm R}({\rm PREX2})$ reproduce the data ~\cite{INGEMARSSON20003,BONIN1985381} at 29.3, 40.975, 48.1~MeV/u. 
In our previous paper~\cite{Matsuzaki:2021hdm}, the data at 29.3, 40.975, 48.1~MeV/u yield  
$r_{\rm skin}^{208}=$ 0.416$\pm$0.146 fm. 
Nevertheless, the value is larger than $r_{\rm skin}^{208}({\rm PREX2})$. 
The central values of the data should decrease as $E_{\rm lab}$ tends to zero in the energy range of $E_{\rm lab} \lsim 41$~Mev because of the Coulomb barrier, since $\sigma_{\rm R}({\rm PREX2})$ has such a $E_{\rm lab}$ dependence. 
However, the data at at 29.3~MeV/u is larger than that that at 40.975~Mev/u. We should neglect the data at 29.3~MeV/u. 

Scaling the neutron PREX2 density with 
the ESP-f method to the data at 40.975, 48.1~MeV/u, 
we obtain $r_{\rm skin}^{208}=0.241 \pm 0.304 $~fm that is 
consistent with  $r_{\rm skin}^{208}({\rm PREX2})$. 

%%%%%%%%%%%%%%%%%%%%%%%
%%%  Figure
%%%%%%%%%%%%%%%%%%%%%%%

As for $^{4}$He, the gaussian expansion method (GEM)~\cite{Kameyama:1989zz} as ab initio calculations is applied for $^{4}$He in which 
the Argonne V8' NN interaction (AV8) and the phenomenological three-nucleon interaction are used~\cite{Hiyama:2004nf}. The NNN interaction is adjusted so as to reproduce the binding energies of $^{4}$He. 
As for $^{4}$He, the matter density of the ground state is shown in Fig.~2 of Ref.~\cite{Hiyama:2004nf}. 

Using the GEM proton and neutron densities for 
$^{4}$He, we obtain $r_{\rm skin}^{208}=0.264 \pm 0.303$~fm. The central 
value is very close to that  $r_{\rm skin}^{208}({\rm PREX2})$. 

\subsection{Reanalyses for $^{48}$Ca} 

\subsubsection{Comparison between D1S and D1M for $^{48}$Ca} 
\label{Comparison between D1S and D1M}

\begin{figure}[H]
\begin{center}
 \includegraphics[width=0.45\textwidth,clip]{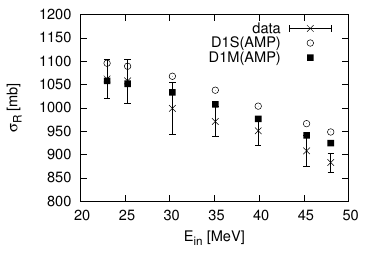}
 \caption{ 
 $E_{\rm lab}=E_{\rm in}$ dependence of reaction cross sections $\sigma_{\rm R}$ 
 for $p$+$^{48}$Ca scattering in $E_{\rm lab}=23 \text{--} 48$~MeV. 
 Circles denote results of the  D1S-GHFB+AMP densities, and 
 squares correspond to those of the  D1M-GHFB+AMP densities
  The data (crosses) are taken from Ref.~\cite{Carlson:1994fq}. 
   }
 \label{Fig-Xsec-$p$+Ca48}
\end{center}
\end{figure}

\begin{figure}[H]
\begin{center}
 \includegraphics[width=0.45\textwidth,clip]{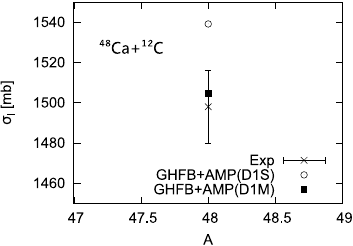}
 \caption{ 
Interaction cross sections $\sigma_{\rm I}$ 
 for $^{48}$Ca+$^{12}$C scattering at 280~MeV/nucleon. 
 Circles denote results of the  D1S-GHFB+AMP densities, and 
 squares correspond to those of the  D1M-GHFB+AMP densities
  The data (crosses) are taken from Ref.~~\cite{Tanaka:2019pdo}. 
   }
 \label{Fig-Xsec-Ca48+C12}
\end{center}
\end{figure}

 Figure~ \ref{Fig-Xsec-$p$+Ca48} shows  $\sigma_{\rm R}$ 
as a function of $E_{\rm lab}$ for $p$+$^{48}$Ca scattering. 
The results of the  D1M-GHFB+AMP densities yield better agreement with 
the data~\cite{Carlson:1994fq} than those of the  D1S-GHFB+AMP densities. 
This is true for $^{48}$Ca+$^{12}$C scattering at 280~MeV/nuleon~\cite{Tanaka:2019pdo}, as shown in Fig.~ \ref{Fig-Xsec-Ca48+C12}.

%%%%%%%%%%%%%%%%%%%%%%%
%%%  Figure
%%%%%%%%%%%%%%%%%%%%%%%

%%%%%%%%%%%%%

%%%%%%%%%%%%%%%%%%%%%%%
%%%  Figure
%%%%%%%%%%%%%%%%%%%%%%%

D1M is thus better than D1S for both $p$+$^{48}$Ca and $^{48}$Ca+$^{12}$C scattering.

\subsubsection{$^{48}$Ca+$^{12}$C  scattering  
in $E_{\rm lab}=280$~MeV/u}

Figure~\ref{Fig-Xsec-Ca48+C12} shows $\sigma_{\rm I}$ 
 for $^{48}$Ca+$^{12}$C scattering at 280~MeV/nucleon.
Scaling the D1M-GHFB+AMP proton and neutron densities for $^{48}$Ca , we can obtain 
\bea
r_{\rm skin}^{48}({\rm skin})=0.180 \pm	0.058~{\rm fm}. 
\eea
Since we do not adopt any fine-tuning factor, we use the resulting values 
of Table~ \ref{TW-Ca48+C} as reference values for $^{48}$Ca.  
We can obtain $r_{\rm m}({\rm CREX})$  and $r_{\rm n}({\rm CREX})$ 
from the CREX value of Eq.~\eqref{CREX-value} and $r_{\rm p}({\rm exp})=3.385~{\rm fm}$~\cite{Angeli:2013epw} of electron scattering.
Our value on $r_{\rm m}({\rm exp})$ is slightly larger than those of CREX. 
As shown in Table~\ref{TW-Ca48+C}, the $r_{\rm n}$ of D1M-GHFB+AMP is very close to the result $r_{\rm n}({\rm ref})$ 
determined from $^{48}$Ca+$^{12}$C scattering.

%\squeezetable
%%%%%%%%%%%%%%
\begin{table}[htb]
\begin{center}
\caption
{Values of   $r_{\rm m}({\rm exp})$,  $r_{\rm n}({\rm exp})$, 
$r_{\rm skin}({\rm exp})$ 
together with $r_{\rm p}({\rm exp})$ 
deduced from the electron scattering~\cite{Angeli:2013epw}. 
The radii are shown in units of fm.  
 }
\begin{tabular}{cccccc}
\hline\hline
 & Ref. & $r_{\rm p}({\rm exp})$ & $r_{\rm m}({\rm exp})$ &  $r_{\rm n}({\rm exp})$ & $r_{\rm skin}({\rm exp})$ \\
\hline
 $^{48}$Ca & CREX & $3.385$ & $3.456 \pm 0.030$ & $3.506 \pm 0.050$ & $0.121 \pm 0.050$ \\
 $^{48}$Ca & ref   & $3.385$ & $3.491 	\pm 0.035$ & 
 $3.565 \pm 0.058$ & $0.180 \pm 0.058$ \\
 $^{48}$Ca & D1M   & $3.417$ & $3.504$ & 
 $3.564 $ & $0.147$ \\
\hline
\end{tabular}
 \label{TW-Ca48+C}
 \end{center} 
 \end{table}
%%%%%%%%%%%%%% 

%%%%%%%%%%%%%

%%%%%%%%%%%%%

\subsubsection{$p$+$^{48}$Ca scattering  in $E_{\rm lab}=23 \text{--} 48$~MeV}
\label{sec:48Ca}

\begin{figure}[H]
\begin{center}
 \includegraphics[width=0.5\textwidth,clip]{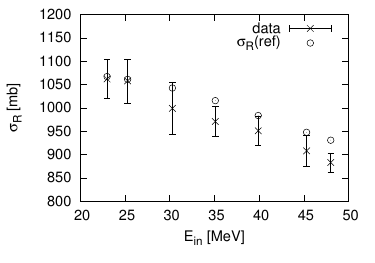}
 \caption{ 
Reaction cross sections $\sigma_{\rm R}$ 
as a function of $E_{\rm lab}=E_{\rm in}$ for $p$+$^{48}$Ca scattering in $E_{\rm lab}=23 \text{--} 48$~MeV.
 Open circles denote results of $\sigma_{\rm R}({\rm ref})$.
  The data (crosses) are taken from Ref.~\cite{Carlson:1994fq}. 
   }
 \label{Fig-RXsec-$p$+Ca48}
\end{center}
\end{figure}

The $\sigma_{\rm R}({\rm ref})$  calculated with 
the Kyushu $g$-matrix folding model with the proton and neutron densities having 
$r_{\rm p}({\rm ref})$ and $r_{\rm n}({\rm ref})$ are compared with  
the data~\cite{Carlson:1994fq} in Fig.~ \ref{Fig-RXsec-$p$+Ca48}. 
The fine-tuning factor $f$ is obtained by averaging  $\sigma_{\rm R}({\rm exp})/\sigma_{\rm R}({\rm ref})$ over $E_{\rm lab}$. The resulting value is 
$f=0.968537$. 
The $f~\sigma_{\rm R}({\rm ref})$ are scaled so as to reproduce  the data~\cite{Carlson:1994fq}. This procedure is nothing but ESP-F. 

%%%%%%%%%%%%%%%%%%%%%%%
%%%  Figure
%%%%%%%%%%%%%%%%%%%%%%%

%%%%%%%%%%%%%

The resulting skin value is 
\bea
r_{\rm skin}^{48}({\rm skin})=0.163 \pm 0.037~{\rm fm}. 
\eea
The results are tabulated in Table~\ref{reference values}. 
Our skin value is consistent with  $r_{\rm skin}^{48}({\rm CREX})$, 
as shown in Table~\ref{reference values}. 

%\squeezetable
%%%%%%%%%%%%%%
\begin{table}[htb]
\begin{center}
\caption
{Values of   $r_{\rm m}({\rm exp})$,  $r_{\rm n}({\rm exp})$, 
$r_{\rm skin}({\rm exp})$ 
together with $r_{\rm p}({\rm exp})$ 
deduced from the electron scattering~\cite{Angeli:2013epw}. 
The radii are shown in units of fm.  
 }
\begin{tabular}{cccccc}
\hline\hline
 & Ref. & $r_{\rm p}({\rm exp})$ & $r_{\rm m}({\rm exp})$ &  $r_{\rm n}({\rm exp})$ & $r_{\rm skin}({\rm exp})$ \\
\hline
 $^{48}$Ca & CREX & $3.385$ & $3.456 \pm 0.030$ & $3.506 \pm 0.050$ & $0.121 \pm 0.050$ \\
 $^{48}$Ca & TW                                 & $3.385$ & $3.481 \pm 0.022$ & $3.548 \pm 0.037$ & $0.163 \pm 0.037$ \\
\hline
\end{tabular}
 \label{reference values}
 \end{center} 
 \end{table}
%%%%%%%%%%%%%% 

\subsection{Analyses of $\sigma_{\rm I}$}

Our  $r_{\rm m}(\sigma_{\rm I})$ values calculated with the LF $t$-matrix 
folding model are consistent with those in 
Table 3 (optical limit) of Ref.~\cite{Ozawa:2001hb} for N, O, C isotopes.
As for N isotopes, our $r_{\rm m}(\sigma_{\rm I})$, however,  agree 
with the upper bound of $r_{\rm m}(\sigma_{\rm I})$ of Ref.~\cite{Bagchi:2019sua} for $A=14, 15, 17 \text{--} 22$. The consistency between our skin values $r_{\rm skin}(\sigma_{\rm I})$ and those of previous works~\cite{Bagchi:2019sua,Kaur:2022yoh,Kanungo:2016tmz} are shown blow. 

\subsubsection{N isotopes}

We use  the data $\sigma_{\rm I}$~\cite{Ozawa:2000gx,Ozawa:2001hb,Ozawa:1993ua,Chulkov:1995mh} for  $^{14 \text{--} 23}$N+$^{12}$C scattering  in $710 \text{--} 1020$~MeV/u; 
see Table 1 of Ref. \cite{Ozawa:2001hb} for the values of 
$\sigma_{\rm I}$.

\begin{figure}[H]
\begin{center}
 \includegraphics[width=0.5\textwidth,clip]{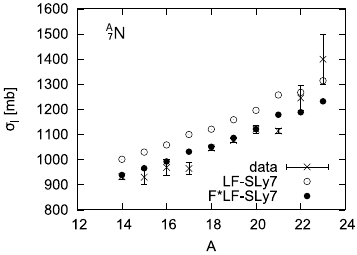}
 \caption{ 
 $A$ dependence of interaction cross sections $\sigma_{\rm I}$ 
 for $^{A}$N+$^{12}$C scattering. 
 Open circles stand for the results of the  LF $t$-matrix folding  model with the  SLy7 densities. 
  Closed circles correspond to  those of $F \sigma_{\rm R}({\rm LF})$. 
 The data are taken from Refs.~\cite{Ozawa:2000gx,Ozawa:2001hb,Ozawa:1993ua,Chulkov:1995mh}; see Table 1 of Ref.~\cite{Ozawa:2001hb}. 
   }
 \label{Fig-RXsec-N+12C-1}
\end{center}
\end{figure}

Figure~\ref{Fig-RXsec-N+12C-1} shows $A$ dependence of interaction cross sections $\sigma_{\rm I}$ for $^{14 \text{--} 23}$N+$^{12}$C scattering, where $A$ is the mass number.
The LF $t$-matrix folding model overshoots $\sigma_{\rm I}$~\cite{Ozawa:2000gx,Ozawa:2001hb,Ozawa:1993ua,Chulkov:1995mh}. 
The renormalized $F \sigma_{\rm R}({\rm LF})$ with $F=0.93766$
reproduces the data~\cite{Ozawa:2000gx,Ozawa:2001hb,Ozawa:1993ua,Chulkov:1995mh}. 
 
%%%%%%%%%%%%%%%%%%%%%%%
%%%  Figure
%%%%%%%%%%%%%%%%%%%%%%%

The SLy7  proton and neutron densities are scaled so that the 
 $F \sigma_{\rm R}({\rm LF})$ can reproduce the data under the condition of  $r_{\rm p,scaling}=r_{\rm p}({\rm exp})$.  

Our results thus obtained  are tabulated in Table  \ref{TW values-N}.
 The $r_{\rm skin}(\sigma_{\rm I})$ of $^{14}$N is close to PREX2 value. 

\begin{table}[H]
\caption
{Values  on  $r_{\rm p}$, $r_{\rm m}$, $r_{\rm n}$, $r_{\rm skin}$ for N isotopes. 
The $r_{\rm p}$ are deduced from 
the charge radii~\cite{Angeli:2013epw} for  $^{14,15}$N and  
the $r_{\rm p}({\rm CC}) $ of  Ref.~\cite{Bagchi:2019sua} are 
used for $A=17 \text{--} 22$. 
 }
 \begin{tabular}{|c|c|c|c|c|c|c|c|c|}
 \hline
$A$ & $r_{\rm skin}$ & error & $r_{\rm m}$ & error & $r_{\rm n}$ & error & $r_{\rm p}$ & error  \cr
 \hline
 \hline
14 &0.259   & 0.079  & 2.553  & 0.042  &2.680  & 0.079  & 2.42041 & 0.000 \cr
15 & 0.091  & 0.175  & 2.523  & 0.096  & 2.565  & 0.175  &2.47402  & 0.000 \cr
17 & 0.010  & 0.182  & 2.556  & 0.079  & 2.560  & 0.152  & 2.55  & 0.03 \cr 
18 &0.368   & 0.106  & 2.761  & 0.038  & 2.898  & 0.076  & 2.53  & 0.03 \cr   
19 & 0.432  & 0.105  & 2.801  & 0.041  & 2.952  & 0.075  & 2.52  & 0.03 \cr
20 & 0.531  & 0.128  & 2.876  & 0.059  & 3.051  & 0.098  & 2.52  & 0.03 \cr
21 & 0.565  & 0.099  & 2.879  & 0.040  & 3.055  & 0.069  & 2.49  & 0.03 \cr 
22 & 0.828  & 0.212  & 3.119  & 0.127  & 3.358  & 0.182  & 2.53  & 0.03 \cr   
16 &   &   &2.608   & 0.227  &  &   &   &  \cr 
23 &   &   &3.415  & 0.220  &   &   &   &  \cr   
 \hline
 \end{tabular}
 \label{TW values-N}
\end{table}
%%%%%%%%%%%%%%%%%%%%%%%%%%%%%%%%%%%%%%%%%%%%%%%%%%%%%%%%%%%%%%%%%%%%%%%%%%

\begin{figure}[H]
\begin{center}
 \includegraphics[width=0.5\textwidth,clip]{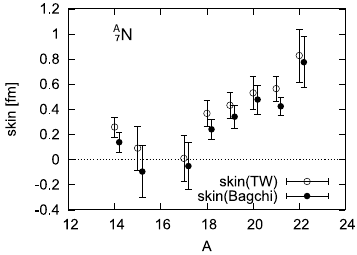}
 \caption{ Comparison between our results and those of  Ref.~\cite{Bagchi:2019sua} for
 $r_{\rm skin}$ for $^{14,15,17 \text{--} 22}$N.
    }
 \label{Fig-N-skins}
\end{center}
\end{figure}

Figure  \ref{Fig-N-skins} shows $A$ dependence of  the $r_{\rm skin}$ 
for $^{14,15,17 \text{--} 22}$N.
Our skin values are compared with those of  Ref.~\cite{Bagchi:2019sua}. 
Our results are slightly larger than theirs. 

%%%%%%%%%%%%%%%%%%%%%%%
%%%  Figure
%%%%%%%%%%%%%%%%%%%%%%%

\subsubsection{O isotopes}

\begin{figure}[H]
\begin{center}
 \includegraphics[width=0.5\textwidth,clip]{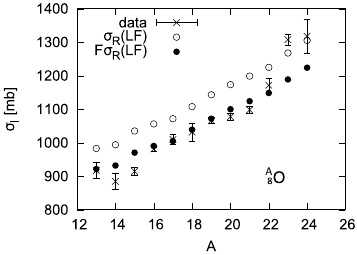}
 \caption{ 
 $A$ dependence of interaction cross sections $\sigma_{\rm I}$ 
 for $^{A}$O+$^{12}$C scattering. 
 Open circles stand for the results of the  LF $t$-matrix folding  model with the D1S-GHFB proton and neutron densities in the 
 spherical limit. 
   Closed circles correspond to  those of 
 $F \sigma_{\rm R}({\rm LF})$. 
 The data are taken from Table 1 of Ref.~\cite{Ozawa:2001hb}. 
   }
 \label{Fig-RXsec-O+12C-1}
\end{center}
\end{figure}

The same procedure is taken for O isotopes. 
As shown in Fig.~\ref{Fig-RXsec-O+12C-1}, 
the LF $t$-matrix folding model overshoots 
the data $\sigma_{\rm I}$ of Refs.~\cite{Ozawa:2000gx,Ozawa:2001hb,Ozawa:1996xsm} in $A=13\text{--}22$. 
The $F \sigma_{\rm R}({\rm LF})$ with $F=0.93766$  
reproduce the data  in $A=13\text{--}22$.

%%%%%%%%%%%%%%%%%%%%%%%
%%%  Figure
%%%%%%%%%%%%%%%%%%%%%%%

The D1S-HFB  proton and neutron densities in 
the spherical limit are scaled so that the $F \sigma_{\rm R}({\rm LF})$ can reproduce the data 
 under the condition of  $r_{\rm p,scaling}=r_{\rm p}({\rm exp})$.  
 
Our results for O isotopes are tabulated in Table  \ref{TW values-O}. 
The skin value is larger for $^{17}$O.

\begin{table}[H]
\caption
{Values  on  $r_{\rm p}$, $r_{\rm m}$, $r_{\rm n}$, $r_{\rm skin}$ for O isotopes. 
The $r_{\rm p}$ are based on the charge radii~\cite{Angeli:2013epw} 
for  $^{16,17,18}$O and  the $r_{\rm p}({\rm CC}) $ 
of  Ref.~\cite{Kaur:2022yoh} for $A=19 \text{--} 24$. 
 }
 \begin{tabular}{|c|c|c|c|c|c|c|c|c|}
 \hline
$A$ & $r_{\rm skin}$ & error & $r_{\rm m}$ & error & $r_{\rm n}$ & error & $r_{\rm p}$ & error  \cr
 \hline
 \hline
16 &0.027   & 0.030  & 2.582  & 0.015  &2.596  & 0.030  & 2.569 & 0.000 \cr
17 & 0.197  & 0.067  & 2.672  & 0.037  & 2.763 & 0.067  &2.566  & 0.000 \cr
18 &0.059  & 0.111  & 2.684  & 0.063  & 2.710  & 0.111  & 2.651  & 0.000 \cr 
19 &0.322   & 0.088  & 2.741 & 0.024  & 2.872  & 0.058  & 2.55  & 0.03 \cr   
20 & 0.409  & 0.072  & 2.783  & 0.026  & 2.939 & 0.052  & 2.53  & 0.02 \cr
21 & 0.380  & 0.070  & 2.771  & 0.026  & 2.910  & 0.050  & 2.53  & 0.02 \cr 
22 & 0.633  & 0.101  & 2.919  & 0.049  & 3.133  & 0.081  & 2.50  & 0.02 \cr   
23 & 0.895  & 0.089  & 3.192  & 0.034  & 3.475  & 0.059  & 2.58  & 0.03 \cr
24 & 0.975  & 0.202  & 3.193 & 0.108  & 3.485  & 0.162  & 2.51  & 0.04 \cr
13 &   &   &2.532   & 0.063  &  &   &   &  \cr
14 &   &   &2.401   & 0.061  &  &   &   &  \cr
15 &   &   &2.420   & 0.034  &  &   &   &  \cr 
 \hline
 \end{tabular}
 \label{TW values-O}
\end{table}
%%%%%%%%%%%%%%%%%%%%%%%%%%%%%%%%%%%%%%%%%%%%%%%%%%%%%%%%%%%%%%%%%%%%%%%%%%

\begin{figure}[H]
\begin{center}
 \includegraphics[width=0.45\textwidth,clip]{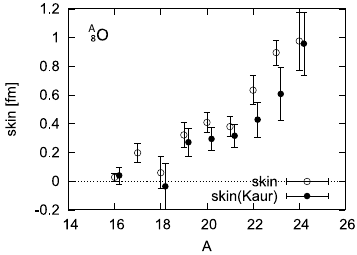}
 \caption{ Comparison of $r_{\rm skin}$ between ours and those of Ref.~\cite{Kaur:2022yoh} for O isotopes. 
    }
 \label{Fig-skins}
\end{center}
\end{figure}

As for O isotopes, our $r_{\rm skin}$ values are slightly 
larger than those of Ref.~\cite{Kaur:2022yoh} for $A=16, 18 \text{--} 24$. As for $A=13 \text{--} 24$, as shown in Fig.~\ref{Fig-skins}. 

%%%%%%%%%%%%%%%%%%%%%%%
%%%  Figure
%%%%%%%%%%%%%%%%%%%%%%%

\subsubsection{C isotopes}

As for  $^{12}$C+$^{12}$C scattering, 
the $F \sigma_{\rm R}({\rm LF})$ reproduces 
the data~\cite{Tanihata:1988ub,Ozawa:2001hb} at 
790, 950~MeV/u within error-bars, as shown 
in Fig.~ \ref{Fig-RXsec-12C+12C}. 
The phenomenological proton and neutron densities are scaled 
so as to reproduce the data under the condition of  
$r_{\rm p,scaling}=r_{\rm p}({\rm exp})$.  
The average of two $r_{\rm m}$ values is taken.  

The projectile $^{12}$C densities should be the 
same as the target $^{12}$C ones. 
The reason why we use 
the phenomenological densities are that 
the $r_{\rm m}({\rm th})$ of the 
phenomenological matter densities 
yield better agreement with the experimental values of Table  \ref{reference values-0} than  that of 
the D1S-GHFB+AMP matter densities

The same procedure is taken for  $^{14,16,18}$C+$^{12}$C scattering 
in which the D1S-GHFB+AMP proton and neutron densities 
for  $^{14,16,18}$C. 

Our results for C isotopes are tabulated in Table  \ref{TW values-C}.

\begin{table}[H]
\caption
{Values  on  $r_{\rm p}$, $r_{\rm m}$, $r_{\rm n}$, $r_{\rm skin}$ for C isotopes. 
The $r_{\rm p}$ are deduced from 
the charge radii~\cite{Angeli:2013epw} for  $^{12,14}$C and  
$r_{\rm p}({\rm CC}) $ of  Ref.~\cite{Kanungo:2016tmz} are used for $A=16,18$. 
 }
 \begin{tabular}{|c|c|c|c|c|c|c|c|c|}
 \hline
$A$ & $r_{\rm skin}$ & error & $r_{\rm m}$ & error & $r_{\rm n}$ & error & $r_{\rm p}$ & error  \cr
 \hline
 \hline
12 & 0.058  & 0.031  & 2.356  & 0.016  & 2.385  & 0.031  & 2.327  & 0.000 \cr
14 & 0.079  & 0.087  & 2.415  & 0.051  & 2.449  & 0.088  & 2.370  & 0.000 \cr
16 & 0.586  & 0.099  & 2.781  & 0.027  & 2.986  & 0.059  & 2.400  & 0.040 \cr 
18 & 0.733  & 0.104  & 2.900  & 0.035  & 3.123  & 0.064  & 2.390  & 0.040 \cr   
 \hline
 \end{tabular}
 \label{TW values-C}
\end{table}
%%%%%%%%%%%%%%%%%%%%%%%%%%%%%%%%%%%%%%%%%%%%%%%%%%%%%%%%%%%%%%%%%%%%%%%%%%

\begin{figure}[H]
\begin{center}
 \includegraphics[width=0.45\textwidth,clip]{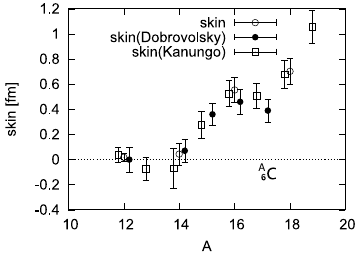}
 \caption{ Comparison of $r_{\rm skin}$ between ours and those of  Refs.~\cite{Dobrovolsky:2021ggf,Kanungo:2016tmz} for C isotopes. 
    }
 \label{Fig-skins-C}
\end{center}
\end{figure}

Figure  \ref{Fig-skins-C} shows $A$ dependence of $r_{\rm skin}$
for $^{12 \text{--} 18}$C.
Our $r_{\rm skin}$ values are consistent with those of Refs.~\cite{Dobrovolsky:2021ggf,Kanungo:2016tmz}.  

%%%%%%%%%%%%%%%%%%%%%%%
%%%  Figure
%%%%%%%%%%%%%%%%%%%%%%%

\subsubsection{Shell effects}

We analyzed the data $\sigma_{\rm I}$~\cite{TAKECHI2021104923} 
on $^{42-51}$Ca+$^{12}$C scattering 
by using Kyushu (chiral) $g$-folding model 
with D1S-GHFB  
proton and neutron densities with and without AMP.  

\begin{figure}[H]
\begin{center}
 \includegraphics[width=0.45\textwidth,clip]{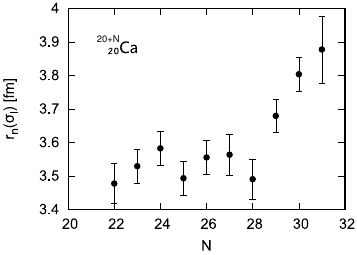}
 \caption{ $N$ dependence of $r_{\rm n}$ for $^{42 \text{--} 51}$Ca.
    }
 \label{Fig-Rn}
\end{center}
\end{figure}

Figure  \ref{Fig-Rn} shows our $r_{\rm n}$ 
of Ref.~\cite{TAKECHI2021104923}  values 
as a function of $N$ for $^{42 \text{--} 51}$Ca.
Our $r_{\rm n}$ values are minimized at $N=28$. 
This is the fact that $N=28$ is a major shell. 

%%%%%%%%%%%%%%%%%%%%%%%
%%%  Figure
%%%%%%%%%%%%%%%%%%%%%%%

\begin{figure}[H]
\begin{center}
 \includegraphics[width=0.45\textwidth,clip]{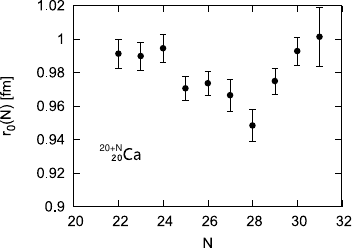}% PDF
 \caption{ $N$ dependence of  $r_0(N)=r_{\rm m}(N)/A^{1/3}$ for $^{42 \text{--} 51}$Ca.
    }
 \label{Fig-R0-Ca}
\end{center}
\end{figure}

Figure  \ref{Fig-R0-Ca} shows our $r_0(N)=r_{\rm m}(N)/A^{1/3}$ values 
as a function of $N$ for $^{42 \text{--} 51}$Ca.
Our  $r_0(N)$ values also have a dip in $N=28$. 
 The $r_0(N)$ are useful to find a major shell.

%%%%%%%%%%%%%%%%%%%%%%%
%%%  Figure
%%%%%%%%%%%%%%%%%%%%%%%

Now, we consider the case of O, N, C isotopes by using  $r_0(N)$. 

\begin{figure}[H]
\begin{center}
  \includegraphics[width=0.45\textwidth,clip]{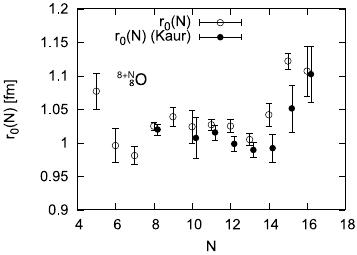}
  \includegraphics[width=0.45\textwidth,clip]{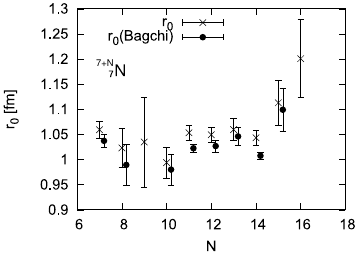}
  \includegraphics[width=0.45\textwidth,clip]{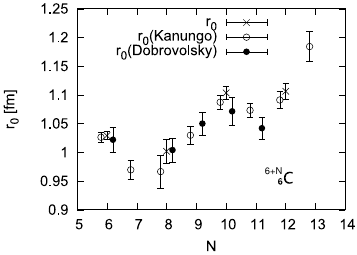} 
 \caption{ Comparison of $r_0(N)=r_{\rm m}(N)/A^{1/3}$ between ours and those of  Ref.~\cite{Kaur:2022yoh} for O isotopes, 
of  Ref.~\cite{Bagchi:2019sua} for N isotopes, of  Refs.~\cite{Kanungo:2016tmz,Dobrovolsky:2021ggf} for C isotopes. 
    }
 \label{Fig-r0-O}
\end{center}
\end{figure}

Figure  \ref{Fig-r0-O} shows $N$ dependence of  
$r_0(N)=r_{\rm m}(N)/A^{1/3}$ for O,N, C isotopes. 
The $r_0(N)$ are minimized at  $N=14$ for N isotopes, This indicates the fact that $N=14$ is a sub-shell. 
The $r_0(N)$ are minimized at  $N=8$ for N, C isotopes. 
This shows the fact that $N=8$ is a major-shell,

%%%%%%%%%%%%%%%%%%%%%%%
%%%  Figure
%%%%%%%%%%%%%%%%%%%%%%%

 \subsubsection{Relation between $r_{\rm m}$ and  total binding energy for  N isotopes}

As for N isotopes, the data on $\beta \equiv r_{\rm m}E_{\rm B}/(A\hbar c)$ hardly depend on $A$ 
for $A=14 \text{--} 22$; note that $E_{\rm B}/A$ is the binding energy per nucleon. 
In fact, the deviation  of  $\beta$ is  much smaller  than the average value; namely, 
\bea
\beta=0.0977 (6) 
\label{Eq-a}
\eea
for $A=14 \text{--} 22$. This indicates that  $r_{\rm m}$ is inversely  proportion to $E_{\rm B}/A$. 

\subsection{ Analyses of $\sigma_{\rm R}$ based on p scattering}

Our $r_{\rm m}(\sigma_{\rm R})$ and $r_{\rm skin}(\sigma_{\rm R})$ calculated with the Kyushu $g$-matrix folding model are shown blow. 
As for N isotopes, the $r_{\rm m}({\rm th})$ calculated with SLy7 
agree with those with D1S-GHFB within 0.2~\%.

\subsubsection{p+$^{14}$N scattering}

We extract  $r_{\rm m}(\sigma_{\rm R})$ from  the data~\cite{CARLSON198557} 
$\sigma_{\rm R}({\rm exp})$ for 
p+$^{14}$N scattering, using the Kyushu $g$-matrix folding model with the Sly7 proton and neutron densities.

\begin{figure}[H]
\begin{center}
 \includegraphics[width=0.5\textwidth,clip]{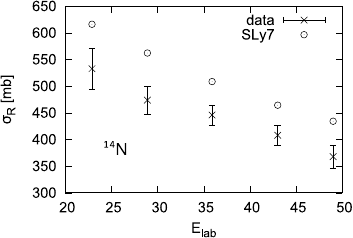}
 \caption{$E_{\rm lab}$ dependence of $\sigma_{\rm R}$ 
 for p+$^{14}$N scattering.  
 Open circles stand for results of the Kyushu $g$-matrix folding  model with the SLy7 densities. 
 The data is taken from Refs.~\cite{CARLSON198557}. 
    }
 \label{Fig-Xsec-p+N14}
\end{center}
\end{figure}

Figure \ref{Fig-Xsec-p+N14} shows our values on $\sigma_{\rm R}$ 
and  the data~\cite{CARLSON198557}. 
The  Kyushu $g$-matrix folding  model with the SLy7 densities almost agrees with the data~\cite{CARLSON198557}. We then introduce     
a fine-tuning factor $f$. 
We use the ESP-f of Ref.~\cite{Wakasa:2022ite} in order to determine $f$.
The fine-tuning factor $f$ is obtained by averaging  $\sigma_{\rm R}({\rm exp})/\sigma_{\rm R}({\rm th})$ over $E_{\rm lab}$. The resulting value is $f=0.86196$. 
 we obtain $r_{\rm m}(\sigma_{\rm R})$, as shown in Table  \ref{TW values-N14}.
The $r_{\rm m}(\sigma_{\rm R})$  is larger than 
$r_{\rm m}(\sigma_{\rm I})$ shown in  Table~\ref{TW values-N14}. 
 The $r_{\rm skin}(\sigma_{\rm R})$ of $^{14}$N is even close to PREX2 value.

%%%%%%%%%%%%%%%%%%%%%%%
%%%  Figure
%%%%%%%%%%%%%%%%%%%%%%%

\begin{table}[hbtp]
\caption
{Values  on $r_{\rm m}(\sigma_{\rm R})$, $r_{\rm p}$,$r_{\rm n}$, $r_{\rm skin}$  for $^{14}$N based on proton scattering.
 }
 \begin{tabular}{|c|c|c|c|c|c|c|c|}
 \hline
$A$ & $r_{\rm skin}(\sigma_{\rm R})$ & error & $r_{\rm m}(\sigma_{\rm R})$ & error & $r_{\rm n}$ & error & $r_{\rm p}$  \cr
 \hline
 \hline
14 & 0.267   & 0.056  & 2.558  &	0.029  & 2.688    & 0.056  & 2.42041 \cr   
 \hline
 \end{tabular}
 \label{TW values-N14}
\end{table}
%%%%%%%%%%%%%%%%%%%%%%%%%%%%%%%%%%%%%%%%%%%%%%%%%%%%%%%%%%%%%%%%%%%%%%%%%%

\subsubsection{p+$^{16}$O scattering}

In Ref.~\cite{INGEMARSSON1999341}, the $\sigma_{\rm R}$ 
have been measured for $^{12}$C, $^{16}$O targets 
at 65.5 MeV. We first derive  $\sigma_{\rm R}({\rm th})$ with the Kyushu $g$-matrix folding model with the phenomenological proton and neutron densities of  $^{12}$C   and introduce a fine-tuning factor $f$ as $f=\sigma_{\rm R}({\rm exp})/\sigma_{\rm R}({\rm th})=0.92449$. 
The reason why we take the phenomenological proton and neutron densities is 
that the phenomenological matter radius $r_{\rm m}(\sigma_{\rm ph})$ of $^{12}$O, $^{16}$O are close to the corresponding experimental 
values, as shown in  Table~\ref{reference values-0} and Table ~\ref{reference valuWS-16}.  
Applying the $f$ value to p+$^{16}$O scattering at 65.5~MeV and 
scaling the the phenomenological densities of  $^{16}$O, 
we obtain  $r_{\rm m}(\sigma_{\rm R})=2.584 \pm 0.053$~fm 
for $^{16}$O. The value is consistent with $r_{\rm m}(\sigma_{\rm I})$, 
as shown in Table ~\ref{reference valuWS-16}.

%\squeezetable
%%%%%%%%%%%%%%
\begin{table}[htb]
\begin{center}
\caption
{Values of   $r_{\rm m}(\sigma_{\rm R})$ and $r_{\rm m}(\sigma_{\rm I})$. 
The values $r_{\rm m}(\sigma_{\rm I})$ are taken from 
 Ref.~\cite{Ozawa:2001hb}. 
 The radii are shown in units of fm.  
 }
\begin{tabular}{cccccc}
\hline\hline
 & $r_{\rm m}(\sigma_{\rm R})$ & $r_{\rm m}(\sigma_{\rm ph})$ & 
 $r_{\rm m}(\sigma_{\rm I})$ &  & \\
\hline
$^{12}$C &  &2.3375
&  & \\
$^{16}$O & $2.584 	\pm 0.053$ & 2.5914
 & $2.54 \pm 0.02$ & \\
\hline
\end{tabular}
 \label{reference valuWS-16}
 \end{center} 
 \end{table}
%%%%%%%%%%%%%% 

\subsection{Halo nature}

Tanihata {\it et al.} determined $r_{\rm m}$ for $^{6-9,11}$Li, $^{12}$C from 
$\sigma_{\rm I}$ at 790~MeV/u~\cite{Tanihata:1988ub} and found that $^{11}$Li is a halo nucleus; see Refs.~\cite{Tanihata:1988ub,Ozawa:2001hb} for the $r_{\rm m}$.

We extracted $r_{\rm skin}$ for $^{6,8}$He by using the LF $t$-matrix 
folding model.  
The nature of  halo is defined only qualitatively. Meanwhile, $r_{\rm skin}$ is defined quantitatively: 
For halo nuclei, 
the $r_{\rm skin}$ are $0.778 \pm 0.041$~fm for $^{6}$He, 
$0.975 \pm 0.204$~fm  $^{11}$Li~\cite{Tanihata:1988ub,Angeli:2013epw,Ozawa:2001hb}, 
$0.853 \pm 0.071$~fm for $^{11}$Be~\cite{Ozawa:2000gx,Angeli:2013epw,Ozawa:2001hb}; 
see Ref.  
\footnote{
The value for $^{6}$He is taken from Ref.~\cite{WAKASA2022105329}. 
The value for $^{11}$Li is calculated from 
$r_{\rm m}=3.12 \pm 0.16$~fm ~\cite{Tanihata:1988ub,Ozawa:2001hb} and $r_{\rm p}=2.381$~fm deduced from 
the charge radius~\cite{Angeli:2013epw}, and 
the value for $^{11}$Be is from  $r_{\rm m}=2.91	\pm 0.05$~fm ~\cite{Ozawa:2000gx,Ozawa:2001hb} 
and $r_{\rm p}=2.338$~fm deduced from 
the charge radius~\cite{Angeli:2013epw}. 
} for the derivation. 
In Fig. 8 of Ref. \cite{PhysRevC.85.064613}, we showed $A$ dependence of $r_{\rm skin}$ for Ne isotopes, where $A$ is the mass number.
The $r_{\rm skin} \approx 0.52$~fm for $^{31}$Ne is much larger than those of $^{30,32}$Ne.
These large skin values come from the halo nature. 
This makes it possible to define the halo nature with $r_{\rm skin}$. 
However, for $^{22}$C as a heaviest halo nucleus at the present stage, 
we cannot extract $r_{\rm skin}$, since $r_p$ is unknown.

As for $^{22}$C, a large increase in the $\sigma_{\rm R}$ 
of $^{22}$C+p scattering from 
that of $^{20}$C+p scattering~\cite{PhysRevLett.104.062701} was observed. 
Here $^{21}$C is unbound. 
Adding p to $^{21}$C yields  $^{22}$N that is a weakly-bound nucleus having 
the single-neutron separation energy $s_{\rm n}=1.54$~MeV. This implies that 
$^{22}$N is a halo nucleus described by the $^{21}$N+n 
two-body model. 
The increase of $\sigma_{\rm R}$ or $\sigma_{\rm I}$ is thus important. 
For the case of $^{22}$N, we then define the nature of halo quantitatively as 
\bea
 {\cal H}_1 = 
 \frac{4 \pi r_{\rm m}(^{22}{\rm N})^2-4 \pi r_{\rm m}(^{21}{\rm N})^2}
 { 4 \pi r_{\rm m}(^{22}{\rm N})^2}
 \eea
and
\bea
  {\cal H}_2= \frac{s_{\rm n}(^{22}{\rm N})}{E_{\rm B}(^{22}{\rm N})/A}
\eea
The nature of halo is realized, when 
$ {\cal H}_1$ are large and $ {\cal H}_2$ is small.   
Our results of Table \ref{TW values-N} yield
 $ {\cal H}_1=0.148$ and $ {\cal H}_2=0.241$ for $^{22}$N. 
For $^{22}$C with $s_{\rm 2n}=10$~keV~\cite{PhysRevLett.104.062701,Ozawa:2001hb}, we get 
$ {\cal H}_1=0.695$ and $ {\cal H}_2=0.001$. 
For $^{11}$Li with  $s_{\rm 2n}=0.369$~MeV ~\cite{Tanihata:1988ub}, 
we obtain $ {\cal H}_1=0.447$ and $ {\cal H}_2=0.044$. We consider that 
 $^{22}$N is halo-like.

\subsubsection{Deformation for $^{12,14,16,18}$C}

In Ref.~\cite{LI2021101440}, Li, Luo and Wang compiled the charge radii $R{\rm ch}$ of 236 nuclei  measured by laser spectroscopy experiment, and calculated the uncertainties. From the $R{\rm ch}$ of Mg isotope chain, the new magic number $N = 14$ can be observed. 
They introduced  Eq. (5) for $R{\rm ch}$.
However, C, N, O isotopes are not included in the  236 nuclei. 

Using Eq. (5) of Ref.~\cite{LI2021101440}, we took the shell corrections (SC) for  $^{16,24}$O, since the SC are shown in 
Ref.~\cite{PhysRevC.81.044322}.  
Using the parameter set (Table A) of Ref.~\cite{LI2021101440}, 
we show $A$ dependence of $R{\rm ch}({\rm WS-1})$, $R{\rm ch}({\rm HFB25-1})$, $R{\rm ch}({\rm WS-1+SC})$, where 
the $R{\rm ch}({\rm WS-1+SC})$ include SC but 
the $R{\rm ch}({\rm WS-1})$ and the $R{\rm ch}({\rm HFB25-1})$ do not. 

\begin{figure}[H]
\begin{center}
 \includegraphics[width=0.5\textwidth,clip]{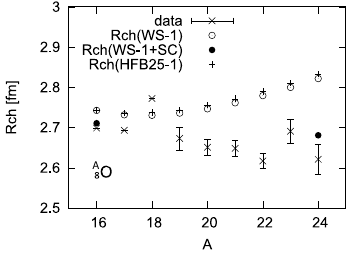}
  \caption{ 
$A$ dependence of $R{\rm ch}({\rm WS-1})$, $R{\rm ch}({\rm HFB25-1})$. 
Open circles denote the $R{\rm ch}({\rm WS-1})$, whereas the symbol ``+'' corresponds to the $R{\rm ch}({\rm HFB25-1})$. 
Closed circles stand for $R{\rm ch}({\rm WS-1+SC})$.
The data (crosses) are taken from Refs.~\cite{Angeli:2013epw,Kaur:2022yoh}. 
   }
 \label{Fig-Rch+O}
\end{center}
\end{figure}

In Fig.~\ref{Fig-Rch+O}, for simplicity, we take the deformation parameters $\b_2=0$ and $\b_4=0$. 
As for  $^{16,17,18}$O, the $R{\rm ch}$ are taken from Ref.~\cite{Angeli:2013epw}. 
As for  $^{19--24}$O, the proton radii of Ref.~\cite{Kaur:2022yoh} are 
transformed into the corresponding $R{\rm ch}$. The data~\cite{Angeli:2013epw,Kaur:2022yoh} on 
$R{\rm ch}$ are compared with the $R{\rm ch}({\rm WS-1})$, 
the $R{\rm ch}({\rm HFB25-1})$, the $R{\rm ch}({\rm WS-1+SC})$. 
As for $^{16, 24}$O, the $R{\rm ch}({\rm WS-1+SC})$ are larger 
than the data. When $\b_2>0$, the $R{\rm ch}({\rm WS-1+SC})$ increase.
This fact indicates that one should not use Eq. (5) of Ref.~\cite{LI2021101440}.

%%%%%%%%%%%%%%%%%%%%%%%
%%%  Figure
%%%%%%%%%%%%%%%%%%%%%%%

SLy7 and D1S show that N, O isotopes are spherical. We then consider deformation on C isotopes.  
D1S-GHFB+AMP yields $\b_2=$
-0.378, -0.211, -0.307, -0.345 for $^{12,14,16,18}$C, respectively. 
There is the famous equation on deformation:
\bea
r_{\rm m}^{2}=r_{\rm m,0}^{2} \lfloor 1+ \frac{5\b_2^2}{4\pi} \rfloor ,
\eea
where $r_{\rm m,0}$ is the matter radius in the spherical limit. 
Using the equation, we can extract $r_{\rm m,0}$ from $r_{\rm m}$. 

Figure \label{Fig-Rm+C} shows $r_{\rm m,0}$ and the data 
(Table \ref{TW values-C}) 
on $r_{\rm m}$ 
for $^{12,14,16,18}$C. Deformation effects for the $r_{\rm m}$ are about 14.8\% for $^{12,14,16,18}$C.

%%%%%%%%%%%%%%%%%%%%%%%
%%%  Figure
%%%%%%%%%%%%%%%%%%%%%%%
\begin{figure}[H]
\begin{center}
 \includegraphics[width=0.5\textwidth,clip]{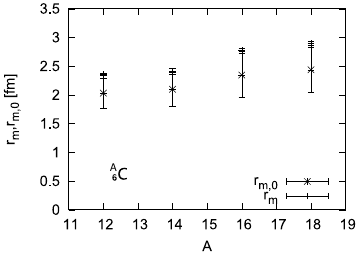}
  \caption{ 
$A$ dependence of $r_{\rm m}$, $r_{\rm m,0}$. 
The  $r_{\rm m,0}$ have errorbars with `'*'', whereas  
the  $r_{\rm m}$ correspond to errorbars with `'-''. 
   }
 \label{Fig-Rm+C}
\end{center}
\end{figure}

\subsubsection{D1S and SLy7 for $r{\rm skin}$
on N isotopes}

\begin{figure}[H]
\begin{center}
 \includegraphics[width=0.5\textwidth,clip]{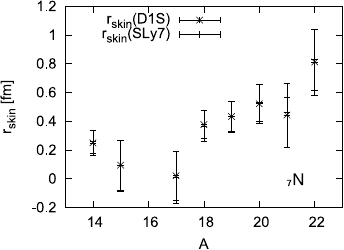}
  \caption{ 
$A$ dependence of $r_{\rm skin}({\rm SLy7})$, 
$r_{\rm skin}({\rm D1S})$.
The $r_{\rm skin}({\rm D1S})$ have errorbars with `'*'', whereas  
the  $r{\rm m}$ correspond to errorbars with `'-''. 
   }
 \label{Fig-skin+N}
\end{center}
\end{figure}

The $r_{\rm skin}({\rm SLy7})$ calculated with SLy7 are compared with 
the $r_{\rm skin}({\rm D1S})$ with D1S in  Fig.~\ref{Fig-skin+N}. 
The $r_{\rm skin}({\rm SLy7})$ are almost the same as $r_{\rm skin}({\rm D1S})$ except for for $^{21}$N. 
The difference between $r_{\rm skin}({\rm SLy7})$ and $r_{\rm skin}({\rm D1S})$ is appreciable only for $^{21}$N. 
In fact, the $r_{\rm skin}({\rm D1S})$ is 
$0.443 \pm	0.221$~fm, whereas  $r_{\rm skin}({\rm SLy7})= 0.565 \pm 0.099$~fm. 
As for  $^{208}$Pb, the difference is very small~\cite{Matsuzaki:2021hdm}.

%%%%%%%%%%%%%%%%%%%%%%%
%%%  Figure
%%%%%%%%%%%%%%%%%%%%%%%

%Results
\section{Discussions for model dependence}
\label{Discussions} 

We use the Kyushu $g$-matrix folding model with no fine-tuning
factor ($F=1$) in $E_{\rm lab} \lsim 410$ MeV/u and the LF $t$-matrix folding model with the fine-tuning $F$ in $E_{\rm lab} \gsim 410$ MeV/u.
The model dependence stems from (A) the proton and neutron densities
used and (B) $F$.
In Sec. \ref{Results},   
our errors only come from those of $\sigma_{\rm T}^{\rm exp}$ 
and  $\sigma_{\rm R}^{\rm exp}$. 
We can minimize the errors based on (A) by choosing the EoS yielding  
that the proton and neutron scaling factor being close  to 1. 
In general, D1S and SLy7(SLy4) are good. As for ${}^{48}$, 
D1M is better than D1S, as shown in Sec.~\ref{sec:48Ca}. 
The errors based on (B) are investigated below.   

As for ${}^{12}{\rm C}$+${}^{12}{\rm C}$ 
scattering at $30 \lsim E_{\rm lab} \lsim 950$ MeV/u, 
the $\sigma_{\rm R}$ of the Kyushu $t$-matrix folding  model agree with those of the LF $t$-matrix folding model $F \sigma_R^{\rm LF}$ with $F=0.93766$ at $E_{\rm lab}=410$ MeV/u, as shown in  Fig.~\ref{Fig-RXsec-12C+12C}.
We then used the $F=0.93766$ value  for C, N, O isotopes. 
As an another fitting, we may use $\sigma_{\rm R}^{\rm exp}=F \sigma_R^{\rm LF}$ at 790~MeV/u. The value is $F=0.94096=0.93766\times1.0035$. 
In order to investigate influence of $F$, we increase the $F=0.93766$ value 
by 0.35\%. As for O isotopes, the central values of $r_{\rm m}$ decrease 
by 0.33\%.  

Now we consider coupled-channel effects in the present folding model.
In $E_{\rm lab} \lsim 410$ MeV/u, we use chiral (Kyushu) $g$-matrix folding model with no fine-tuning factor. The chiral $g$-matrix  include approximately coupled-channel effects as 
nuclear-medium effects obtained by solving the Brueckner-Hartree-Fock method, 
since the chiral $g$-matrix include the back-coupling from all the continuum states; see Ref.~\cite{Toyokawa:2017pdd} for the detained explanation. 

Now we consider compound nucleus effects. Compound nucleus effects appear at low incident energies and are a result of multiple NN collisions. The  effects decrease as $E_{\rm lab}$ increases from 2.491~MeV to 
14.137~MeV in Ref.~\cite{PhysRevC.3.576}; note that the compound -nucleus outgoing processes appear only in s-wave. 
We then take $E_{\rm lab}=$ 14.137~MeV 
as an incident neutron-beam energy generated 
from Li(d,n).     
In fact, the central value of $\sigma_{\rm T}({\rm PREX2})$ is very close 
to that of the $\sigma_{\rm T}({\rm exp})$~\cite{PhysRevC.3.576}, as shown in Fig.~\ref{Fig-TXsec-n+Pb}. 
The small difference between  our present value $r_{\rm skin}^{208}=0.309 \pm  0.057$~fm and $r_{\rm skin}^{208}({\rm PREX2})$ may come from compound nucleus effects.

\section{Summary}
\label{Sec:Summary}

Foster {\it et al.} made high-precision measurement for total neutron cross sections  $\sigma_{\rm T}$~\cite{PhysRevC.3.576} of n+$^{208}$Pb scattering at  $E_{\rm lab}= 14.137$~MeV. We extract $r_{\rm skin}^{208}=0.309 \pm  0.057$~fm from the $\sigma_{\rm T}$, using the chiral (Kyushu) $g$-matrix folding model with the D1S-GHFB+AMP proton and neutron densities. The value  agrees with  $r_{\rm skin}^{208}({\rm PREX2})$ and  is consistent with $r_{\rm skin}^{208}=0.278 \pm 0.035$~fm~\cite{Tagami:2020bee} from  $\sigma_{\rm R}$ on p+$^{208}$Pb scattering.

As for $^{48}$Ca, we determine 
$r_{\rm skin}^{48}({\rm skin})=0.163 \pm 0.037~{\rm fm}$ from the 
$\sigma_{\rm R}$~\cite{Carlson:1994fq} on p+$^{48}$Ca scattering, using the Kyushu $g$-matrix folding model with the D1M-GHFB+AMP proton and neutron densities. We show that D1M-GHFB+AMP is better than D1S-GHFB+AMP for the matter radius and the binding  energy. 
Our skin value is consistent with  $r_{\rm skin}^{48}({\rm CREX})$.

As for $^{4}$He, the Gaussian expansion method (GEM)~\cite{Kameyama:1989zz} as ab initio calculations is applied for $^{4}$He~\cite{Hiyama:2004nf}. 
Using the Kyushu $g$-matrix folding model with the D1S-GHFB+AMP proton and neutron densities for $^{208}$Pb and the GEM proton and neutron densities for 
$^{4}$He, we obtain $r_{\rm skin}^{208}=0.264 \pm 0.303$~fm from 
$\sigma_{\rm R}$ of $^{4}$He+$^{208}$Pb scattering 
at 40.975, 48.1~MeV/u.  
The value is very close to that  $r_{\rm skin}^{208}({\rm PREX2})$.

Using the Kyushu $g$-matrix and the LF folding model, among O, N, C isotopes, we find that $r_{\rm skin}= 0.267 \pm 0.056$~fm for $^{14}$N and 
$r_{\rm skin}= 0.197 \pm 0.067$~fm for $^{17}$O as stable nuclei having 
large skin value. The value $r_{\rm skin}= 0.267 \pm 0.056$~fm for $^{14}$N is consistent with  $r_{\rm skin}^{208}({\rm PREX2})$.
The value $r_{\rm skin}= 0.197 \pm 0.067$~fm for $^{17}$O 
is close to   $r_{\rm skin}^{208}({\rm PREX2})$.

As for the nucleus having the property $Z=N$, the $r_{\rm skin}$  
is zero, if the Coulomb interaction is switched off in the Hamiltonian 
for   $^{14}$N. This is because the realistic NN interaction is invariant under the interchange between proton and neutron.  The thick skin value shows that the Coulomb effects are important. 

The $r_0(N)=r_{\rm m}(N)/A^{1/3}$ are useful to find a major and a sub-major shell. When the $N=8$ core is hard, there should be no dip. In fact, there is no dip for 
O isotopes. As for N, C isotopes, there appears a dip at $N=8$, indicating that the $N=8$ core becomes soft. As for O, N, C isotopes, there appears a dip at $N=14$, indicating that the $N=14$ core is a  soft one.

\section*{Acknowledgements}
We thank Kamimura for providing us with his matter densities of $^{4}$He and 
thank Fukui for his contribution.   

\bibliography{Folding-v23}

%%%%%%%%%%%%%%%%%%%%%%%%%%%%%%%%%%%%%%%%%%

\end{document}